\documentclass[12pt]{amsart}
\usepackage{amsmath}
\usepackage{amsxtra}
\usepackage{amscd}
\usepackage{amsthm}
\usepackage{amsfonts}
\usepackage{amssymb}
\usepackage{eucal}

\input prepictex
\input pictex
\input postpictex
\usepackage{mathptm}

\usepackage{verbatim}

\newtheorem{theorem}{Theorem}[section]

\newtheorem{lemma}[theorem]{Lemma}

\newtheorem{proposition}[theorem]{Proposition}

\theoremstyle{definition}
\newtheorem{definition}[theorem]{Definition}

\theoremstyle{remark}
\newtheorem{remark}[theorem]{Remark}

\numberwithin{equation}{section}

\newcommand{\R}{{\mathbb R}}
\newcommand{\C}{{\mathbb C}}

\newcommand{\barh}{{\bar h}}
\newcommand{\cA}{{\mathcal A}}
\newcommand{\cG}{{\mathcal G}}

\newcommand{\cC}{{\mathcal C}}

\newcommand{\cM}{{\mathcal M}}
\newcommand{\cR}{{\mathcal R}}
\newcommand{\cE}{{\mathcal E}}

\newcommand{\bM}{{\mathbf M}}

\newcommand{\lAm}{{_l\cA^m}}
\newcommand{\rAm}{{\cA_r^m}}

\newcommand{\g}{{\mathbf g}}

\begin{document}
\title[Projective modules and embedded noncommutative spaces]
{Projective module description of embedded noncommutative spaces}

\author{R.B. Zhang}
\address{School of Mathematics and Statistics,
University of Sydney, Sydney, Australia}
\email{ruibin.zhang@sydney.edu.au}

\author{Xiao Zhang}
\address{Institute of Mathematics, Academy of Mathematics and Systems Science,
Chinese Academy of Sciences, Beijing, China}
\email{xzhang@amss.ac.cn}

\begin{abstract}
An algebraic formulation is given for the
embedded noncommutative spaces over the Moyal algebra
developed in a geometric framework in \cite{CTZZ}.
We explicitly construct the projective modules corresponding
to the tangent bundles of the embedded noncommutative spaces,
and recover from this algebraic formulation the metric, Levi-Civita
connection and related curvatures,
which were introduced geometrically in \cite{CTZZ}.
Transformation rules for connections and curvatures
under general coordinate changes are given. A bar
involution on the Moyal algebra is discovered, and its consequences
on the noncommutative differential geometry are described.
\end{abstract}
\maketitle

\tableofcontents
\section{Introduction}\label{introduction}

It is a long held belief in physics that the notion of spacetime as
a pseudo Riemannian manifold requires modification at the Planck
scale \cite{Sn, Y}. Theoretical investigations in recent times
strongly supported this view. In particular, the seminal paper
\cite{DFR} by Doplicher, Fredenhagen and Roberts demonstrated
mathematically that coordinates of spacetime became noncommutative
at the Planck scale, thus some form of noncommutative geometry
\cite{Co} appeared to be necessary in order to describe the
structure of spacetime. This prompted intensive activities in
mathematical physics studying various noncommutative generalisations
of Einstein's theory of general relativity \cite{Ch1, Ch2, Mad1, Mad2, Maj, ADMW,
AMV, CTZZ, CT1, CT2, COTZ}. For reviews on earlier works, we refer to
\cite{MH, MH1} and references therein. For more recent developments,
particularly on the study of noncommutative black holes, see
\cite{CT1, CT2, WZZ, R2, Rand2, Rand3, Rand4, Rand5, Rand6}.

In joint work with Chaichian and Tureanu \cite{CTZZ}, we
investigated the noncommutative geometry \cite{Co, GVF} of
noncommutative spaces embedded in higher dimensions. We first
quantised a space by deforming \cite{Ge, Ko} the algebra of
functions to a noncommutative associative algebra known as the Moyal
algebra. Such an algebra naturally incorporates the generalised
spacetime uncertainty relations of \cite{DFR}, capturing key
features expected of spacetime at the Planck scale. We then
systematically investigated the noncommutative geometry of embedded
noncommutative spaces. This was partially motivated by Nash's
isometric embedding theorem \cite{embeddings} and its generalisation
to pseudo-Riemannian manifolds \cite{embeddings-1, embeddings-2, embeddings-3}, which state that
any (pseudo-) Riemannian manifold can be isometrically embedded in
Euclidean or Minkowski spaces. Therefore, in order to study the
geometry of spacetime, it suffices to investigate (pseudo-)
Riemannian manifolds embedded in higher dimensions. Embedded
noncommutative spaces also play a role in the study of branes
embedded in $\R^D$ in the context of Yang-Mills matrix models
\cite{St}.

The theory of \cite{CTZZ} was developed within a geometric framework
analogous to the classical theory of embedded surfaces (see, e.g.,
\cite{DoC}). The present paper further develops the differential
geometry of embedded noncommutative spaces by constructing an
algebraic formulation in terms of projective modules,
a language commonly adopted in noncommutative geometry \cite{Co, GVF}.

We shall first describe the finitely generated projective modules
over a Moyal algebra, which will be regarded as noncommutative
vector bundles on a quantised spacetime. We then construct a
differential geometry of the noncommutative vector bundles,
developing a theory of connections and curvatures on such bundles.
In doing this, we make crucial use of a unique property of the Moyal
algebra, namely, it has a set of mutually commutative derivations
related to the usual partial derivations of functions.

Then we apply the noncommutative differential geometry developed to
study the embedded noncommutative spaces introduced in \cite{CTZZ}.
We explicitly construct the projective modules corresponding to the
tangent bundles of the noncommutative spaces, and recover from this
algebraic formulation the geometric Levi-Civita connections and
related curvatures introduced in \cite{CTZZ}.  This way, the
embedded noncommutative spaces of \cite{CTZZ} acquire a natural
interpretation in the algebraic formalism present here.

Morally one may regard the very definition of a projective module (a
direct summand of a free module) as the geometric equivalent of
embedding a low dimensional manifold isometrically in a higher
dimensional one. In the commutative setting of classical (pseudo-)
Riemannian geometry, we make this connection more precise and
explicit by showing that the projective module description of
tangent bundles studied here is a natural consequence of the
isometric embedding theorems \cite{embeddings, embeddings-1, embeddings-2, embeddings-3}. This
is briefly discussed in Theorem \ref{classical}.

As a concrete example of noncommutative differential geometries over
the Moyal algebra, we study in detail a quantum deformation of a
time slice of the Schwarzschild spacetime. The projection operator
yielding the tangent bundle is given explicitly, and the
corresponding metric is also worked out.

As is well known, one of the fundamental principles of general
relativity is general covariance. It is important to find a
noncommutative version of this principle. By analyzing the
structure of the Moyal algebra, we show that the noncommutative
geometry developed here (initiated in \cite{CTZZ})
retains some notion of ``general covariance".
Properties of the connection and curvature under general coordinate
transformations are described explicitly (see Theorem
\ref{covariance}).

The Moyal algebra (over the real numbers) admits an involution
similar to the bar involution in the context of quantum groups. We
introduce a particularly nice class of noncommutative vector bundles
over the Moyal algebra, which are associated to bar invariant
idempotents and endowed with bar hermitian connections (see Section
\ref{bar}). In this case the bar involution takes the left tangent
bundles to right tangent bundles. We show that the tangent bundles
of embedded noncommutative spaces under a middle condition belong to
this class.

The organisation of the paper is as follows. In Section
\ref{quantisation}, we describe the Moyal algebras and finitely
generated projective modules over them. In Section \ref{bundles} we
discuss the differential geometry of noncommutative vector bundles
on quantum spaces corresponding to Moyal algebras. In Section
\ref{surfaces} we develop the differential geometry of embedded
noncommutative spaces using the language of projective modules.  As
an explicit example,  we study in detail the quantum deformation of
a time slice of the Schwarzschild spacetime in Section
\ref{example}. In Section \ref{transformations} we study the effect
of general coordinate transformations. In Section \ref{bar}, we
investigate properties of noncommutative vector bundles under the
bar involution of the Moyal algebra. Finally, Section
\ref{conclusion} concludes the paper with some general comments and
a discussion of the natural relationship between projective modules
and isometric embeddings in classical (pseudo-) Riemannian geometry.

Before closing this section, we mention that the theory of
\cite{CTZZ} has the advantage of being explicit and easy to use for
computations. Using this theory, we constructed noncommutative
Schwarzschild and Schwarzschild-de Sitter spacetimes in joint work
with Wang \cite{WZZ}. Our long term aim is to develop a theoretical
framework for studying noncommutative general relativity. A variety
of physically motivated methods and techniques were used in the
literature to study corrections to general relativity arising from
the noncommutativity of the Moyal algebra. In particular, references
\cite{ADMW, AMV} studied deformations of the diffeomorphism algebra
as a means for incorporating noncommutative effects of spacetime,
while in \cite{CT1, CT2, COTZ} a gauge theoretical approached was taken.
These approaches differ considerably from the theory of \cite{CTZZ,
WZZ} at the mathematical level.

\section{Moyal algebra and projective modules}\label{quantisation}
We describe the Moyal algebra of smooth functions on an open region
of $\R^n$, and the finitely generated projective modules over the
Moyal algebra. This provides the background material needed in later
sections, and also serves to fix notations.

We take an open region $U$ in $\R^n$ for a fixed $n$, and write the
coordinate of a point $t\in U$ as $(t^1, t^2, \dots, t^n)$. Let
$\barh$ be a real indeterminate, and denote by $\R[[\barh]]$ the
ring of formal power series in $\barh$. Let $\cA$ be the set of formal 
power series in $\barh$ with coefficients being real smooth
functions on $U$. Namely, every element of $\cA$ is of the form
$\sum_{i\ge 0} f_i\barh^i$ where $f_i$ are smooth functions on $U$.
Then $\cA$ is an $\R[[\barh]]$-module in the obvious way.

Fix a constant skew symmetric $n\times n$ matrix $\theta=(\theta_{i
j})$. The Moyal product on $\cA$ corresponding to $\theta$ is a map
\[ \mu: \cA \otimes_{\R[[\barh]]} \cA \longrightarrow \cA,
\quad f\otimes g \mapsto \mu(f, g),
\]
defined by
\begin{eqnarray}\label{multiplication}
\mu(f, g)(t)= \lim_{t'\rightarrow t} \exp^{\barh \sum_{i j}
\theta_{i j}\frac{\partial}{\partial t^i}\frac{\partial}{\partial
t'_j}}f(t) g(t').
\end{eqnarray}
On the right hand side,  $f(t) g(t')$ means the usual product of the
numerical values of the functions $f$ and $g$ at $t$ and $t'$
respectively.

It has been known since the early days of quantum mechanics that the
Moyal product is associative (see, e.g., \cite{Ko} for a reference).
Thus the $\R[[\barh]]$-module $\cA$ equipped with the Moyal product
forms an associative algebra over $\R[[\barh]]$, which is a
deformation of the algebra of smooth functions on $U$ in the sense
of \cite{Ge}. We shall usually denote this associative algebra by
$\cA$, but when it is necessary to make explicit the multiplication,
we shall write it as $(\cA, \mu)$.

The partial derivations $\partial_i:=\frac{\partial}{\partial t^i}$
with respect to the coordinates $t^i$ for $U$ are
$\R[[\barh]]$-linear maps on $\cA$. Since $\theta$ is a constant
matrix, the Leibniz rule is valid. Namely, for any element $f$ and
$g$ of $\cA$, we have
\begin{eqnarray}\label{Leibniz}
\partial_i\mu(f, g)= \mu(\partial_i f, g) + \mu(f, \partial_i g).
\end{eqnarray}
Therefore, the $\partial_i$ ($i=1, 2, \dots, n$) are mutually
commutative derivations of the Moyal algebra $(\cA, \mu)$ on $U$.

\begin{remark}
The usual notation in the literature for $\mu(f, g)$ is $f\ast g$.
This is referred to as the star-product of $f$ and $g$. Hereafter we
shall replace $\mu$ by $\ast$ and simply write $\mu(f, g)$ as $f\ast
g$.
\end{remark}

Following the general philosophy of noncommutative geometry
\cite{Co}, we regard the associative algebra $(\cA, \mu)$ as
defining some {\em quantum deformation of the region $U$}, and
finitely generated projective modules over $\cA$ as (spaces of
sections of) noncommutative vector bundles on the quantum
deformation of $U$ defined by the noncommutative algebra $\cA$. Let
us now briefly describe finitely generated projective modules over
$\cA$.

Given an integer $m>n$, we let $\lAm$ (resp. $\rAm$) be the set of
$m$-tuples with entries in $\cA$ written as rows (respectively, columns). We
shall regard $\lAm$ (respectively, $\rAm$) as a left (respectively, right)
$\cA$-module with the action defined by multiplication from the left
(respectively, right). More explicitly, for $v=\begin{pmatrix}a_1 & a_2 &
\dots & a_m\end{pmatrix}\in\lAm$, and $b\in\cA$, we have $b \ast v =
\begin{pmatrix}b\ast a_1 & b\ast a_2 & \dots & b\ast a_m\end{pmatrix}$.
Similarly for $w=\begin{pmatrix}a_1 \\  a_2 \\ \vdots \\
a_m\end{pmatrix}\in\rAm$, we have $w\ast b = \begin{pmatrix}a_1\ast b \\
a_2\ast b\\ \vdots \\ a_m\ast b\end{pmatrix}$. Let $\bM_m(\cA)$ be
the set of $m\times m$-matrices with entries in $\cA$. We define
matrix multiplication in the usual way but by using the Moyal
product for products of matrix entries, and still denote the
corresponding matrix multiplication by $\ast$. Now for $A=(a_{i j})$
and $B=(b_{i j})$, we have $(A\ast B)=(c_{i j})$ with $c_{i j} =
\sum_{k} a_{i k}\ast b_{k j}$. Then $\bM_m(\cA)$ is an
$\R[[\barh]]$-algebra, which has a natural left (respectively, right) action
on $\rAm$ (respectively, $\lAm$).

A finitely generated projective left (respectively, right) $\cA$-module is
isomorphic to some direct summand of $\lAm$ (respectively, $\rAm$) for some
$m<\infty$. If $e\in \bM_m(\cA)$ satisfies the condition $e\ast
e=e$, that is, it is an idempotent, then
\[
\cM=\lAm\ast e := \{v\ast e \mid v\in \lAm \}, \quad
\tilde\cM=e\ast\rAm := \{e\ast w \mid \in \rAm \}
\]
are respectively projective left and right $\cA$-modules.
Furthermore, every projective left (right) $\cA$-module is
isomorphic to an $\cM$ (respectively, $\tilde\cM$) constructed this way by
using some idempotent $e$.

In Section \ref{surfaces}, we shall give a systematic method for
constructing idempotents (see \eqref{idempotent}).  The
corresponding noncommutative vector bundles include the tangent
bundles of embedded noncommutative spaces introduced in \cite{CTZZ},
which we shall investigate in depth. An explicit example of embedded
noncommutative spaces will be analyzed in detail in Section
\ref{example}. To do this, we need to develop some generalities of
the differential geometry of noncommutative vector bundles using the
language of projective modules over the Moyal algebra.

\section{Differential geometry of noncommutative vector bundles}\label{bundles}

In this section we investigate general aspects of the noncommutative
differential geometry over the Moyal algebra. We shall focus on the
abstract theory here. A large class of examples will be given in
Section \ref{surfaces}, including one which will be worked out in
detail.

As we shall see, the set of mutually commutative derivations
$\partial_i$ ($i=1, 2, \dots, n$) of the Moyal algebra $\cA$ will
play a crucial role in developing the noncommutative differential
geometry.

\subsection{Connections and curvatures}\label{connections}

We start by considering the action of the partial derivations
$\partial_i$ on $\cM$ and $\tilde\cM$. We only treat the left module
in detail, and present the pertinent results for the right module at
the end, since the two cases are similar.

Let us first specify that $\partial_i$ acts on rectangular matrices
with entries in $\cA$ by componentwise differentiation. More
explicitly,
\[
\partial_i B=
\begin{pmatrix} \partial_i b_{1 1} & \partial_i b_{1 2} & \dots &\partial_i b_{1 l}\\
                \partial_i b_{2 1} & \partial_i b_{2 2} & \dots &\partial_i b_{2 l}\\
                  \dots   & \dots    & \dots &\dots\\
                \partial_i b_{k 1} & \partial_i b_{k 2} & \dots &\partial_i b_{k l}
\end{pmatrix} \quad \text{for} \quad B=
\begin{pmatrix}   b_{1 1} & b_{1 2} & \dots &b_{1 l}\\
                  b_{2 1} & b_{2 2} & \dots &b_{2 l}\\
                  \dots   & \dots    & \dots &\dots\\
                  b_{k 1} & b_{k 2} & \dots &b_{k l}
\end{pmatrix}.
\]
In particular, given any $\zeta=v\ast e\in \cM$, where $v\in\lAm$
regarded as a row matrix, we have $\partial_i\zeta = (\partial_i
v)\ast e + v\ast\partial_i(e)$ by the Leibniz rule. While the first
term belongs to $\cM$, the second term does not in general.
Therefore, $\partial_i$ ($i=1, 2, \dots, n$) send $\cM$ to some
subspace of $\lAm$ different from $\cM$.

Let $\omega_i \in \bM_m(\cA)$ ($i=1, 2, \dots, n$) be $m\times
m$-matrices with entries in $\cA$ satisfying the following
condition:
\begin{eqnarray} \label{connect-property}
e\ast\omega_i\ast(1-e) = -e\ast\partial_i e, \quad \forall i.
\end{eqnarray}
Define the $\R[[\barh]]$-linear maps $\nabla_i$ ($i=1, 2, \dots, n$)
from $\cM$ to $\lAm$ by
\[
\nabla_i \zeta = \partial_i\zeta + \zeta \ast\omega_i , \quad
\forall \zeta \in \cM.
\]
Then each $\nabla_i$ is a covariant derivative on the noncommutative
bundle $\cM$ in the sense of Theorem \ref{key} below. They together
define a {\em connection} on $\cM$.
\begin{theorem}\label{key}
The maps $\nabla_i$ ($i=1, 2, \dots, n$) have the following
properties. For all $\zeta\in \cM$ and $a\in \cA$,
\[\nabla_i\zeta \in \cM \quad  \text{and} \quad
\nabla_i(a \ast \zeta)=\partial_i(a)\ast \zeta + a \ast\nabla_i\zeta.\]
\end{theorem}
\begin{proof}
For any $\zeta\in \cM$, we have
\[
\begin{aligned}
\nabla_i(\zeta)\ast e   &=\partial_i(\zeta)\ast e + \zeta \ast\omega_i\ast e\\
                    &= \partial_i\zeta + \zeta \ast(\omega_i\ast e -
                    \partial_i e),
\end{aligned}
\]
where we have used the Leibniz rule and also the fact that $\zeta
\ast e = \zeta$. Using this latter fact again, we have
$ \zeta\ast (\omega_i\ast e - \partial_i e) =  \zeta\ast
(e\ast\omega_i\ast e -e\ast\partial_i e) $, and by the defining
property \eqref{connect-property} of $\omega_i$, we obtain $\zeta
\ast(e\ast\omega_i\ast e -e\ast\partial_i\ast e) = \zeta
\ast\omega_i$. Hence
\[
\nabla_i(\zeta)\ast  e = \partial_i\zeta + \zeta\ast \omega_i
=\nabla_i\zeta,
\]
proving that $\nabla_i\zeta\in\cM$. The second part of the theorem
immediately follows from the Leibniz rule.
\end{proof}

We shall also say that the set of $\omega_i$ ($i=1, 2, \dots, n$) is
a connection on $\cM$. Since $e\ast\partial_i e =
\partial_i(e)\ast(1-e)$, one obvious choice for $\omega_i$ is $\omega_i
= - \partial_i e$, which we shall refer to as the {\em canonical
connection} on $\cM$.

\medskip

By inspecting the defining property \eqref{connect-property} for a
connection, we easily see the following result.
\begin{lemma}
If $\omega_i$ ($i=1, 2, \dots, n$) define a connection on $\cM$,
then so do also $\omega_i + \phi_i\ast e $ ($i=1, 2, \dots, n$) for
any $m\times m$-matrices $\phi_i$ with entries in $\cA$.
\end{lemma}

For a given connection $\omega_i$ ($i=1, 2, \dots, n$), we consider
$[\nabla_i, \nabla_j]= \nabla_i \nabla_j -  \nabla_j \nabla_i$ with
the right hand side understood as composition of maps on $\cM$. By
simple calculations we can show that for all $\zeta\in\cM$,
\[
\begin{aligned}
{[\nabla_i, \nabla_j]}\zeta = \zeta\ast \cR_{i j} \quad \text{with}
\quad \cR_{i j}:=\partial_i \omega_j - \partial_j \omega_i -
[\omega_i, \ \omega_j]_\ast,
\end{aligned}
\]
where $[\omega_i, \omega_j]_\ast = \omega_i\ast \omega_j
-\omega_j\ast \omega_i$ is the commutator. We call $\cR_{i j}$ the
{\em curvature} of $\cM$ associated with the connection $\omega_i$.

For all $\zeta\in\cM$,
\[
\begin{aligned}
{[\nabla_i, \nabla_j]}\nabla_k \zeta &= \partial_k(\zeta) \ast
\cR_{i j}
+ \zeta \ast\omega_k\ast\cR_{i j}, \\
\nabla_k[\nabla_i, \nabla_j] \zeta &= \partial_k(\zeta)\ast \cR_{i
j} + \zeta\ast (\partial_k\cR_{i j} + \cR_{i j}\ast \omega_k).
\end{aligned}
\]
Define the following covariant derivatives of the curvature:
\begin{eqnarray}
\nabla_k \cR_{i j} := \partial_k\cR_{i j} + \cR_{i j}\ast \omega_k -
\omega_k\ast \cR_{i j},
\end{eqnarray}
we have
\[
{[\nabla_k, [\nabla_i, \nabla_j]]} \zeta = \zeta \ast\nabla_k \cR_{i
j}, \quad \forall \zeta\in\cM.
\]
The Jacobian identity $[\nabla_k, [\nabla_i, \nabla_j]]+[\nabla_j,
[\nabla_k, \nabla_i]]+[\nabla_i, [\nabla_j, \nabla_k]]=0$ leads to
\[
\zeta\ast (\nabla_k \cR_{i j}+\nabla_j \cR_{k i}+\nabla_i \cR_{j
k})=0, \quad \forall \zeta\in\cM.
\]
From this we immediately see that $e\ast(\nabla_k \cR_{i j}+\nabla_j
\cR_{k i}+\nabla_i \cR_{j k})=0$. In fact, the following stronger
result holds.
\begin{theorem} The curvature satisfies the following {\em Bianchi identity}:
\[
\nabla_k \cR_{i j}+\nabla_j \cR_{k i}+\nabla_i \cR_{j k}=0.
\]
\end{theorem}
\begin{proof}
The proof is entirely combinatorial. Let
\[
\begin{aligned}
A_{i j k} &= \partial_k\partial_i\omega_j -
\partial_k\partial_j\omega_i, \\
B_{i j k} &= [\partial_i\omega_j, \omega_k]_\ast -
[\partial_j\omega_i, \omega_k]_\ast.
\end{aligned}
\]
Then we can express $\nabla_k\cR_{i j}$ as
\[\nabla_k\cR_{i j} = A_{i j k} + B_{i j k} - \partial_k[\omega_i,\
\omega_j]_\ast - [[\omega_i,\ \omega_j]_\ast, \ \omega_k]_\ast.
\]
Note that
\[
\begin{aligned}
A_{i j k} + A_{j k i} + A_{k i j}&=0, \\
B_{i j k} + B_{j k i} + B_{k i j}&= \partial_k[\omega_i,\
\omega_j]_\ast + \partial_i[\omega_j,\
\omega_k]_\ast+\partial_j[\omega_k,\ \omega_i]_\ast.
\end{aligned}
\]
Using these relations together with the Jacobian identity
\[
[[\omega_i,\ \omega_j]_\ast, \ \omega_k]_\ast+[[\omega_j,\
\omega_k]_\ast, \ \omega_i]_\ast+[[\omega_k,\ \omega_i]_\ast, \
\omega_j]_\ast=0,
\]
we easily prove the Bianchi identity.
\end{proof}

\subsection{Gauge transformations}\label{subsect-gauge}

Let $GL_m(\cA)$ be the group of invertible $m\times m$-matrices with
entries in $\cA$. Let $\cG$ be the subgroup defined by
\begin{eqnarray}\label{gauge-group}
\cG=\{ g\in GL_m(\cA) \mid e\ast g = g\ast e\},
\end{eqnarray}
which will be referred to as the {\em gauge group}. There is a right
action of $\cG$ on $\cM$ defined, for any $\zeta\in \cM$ and
$g\in\cG$, by $\zeta \times g \mapsto \zeta\cdot g:=\zeta \ast g$,
where the right side is defined by matrix multiplication. Clearly,
$\zeta \ast g \ast e = \zeta \ast g$. Hence $\zeta \ast g\in\cM$,
and we indeed have a $\cG$ action on $\cM$.

For a given $g\in\cG$, let
\begin{eqnarray} \label{connect-gauged}
\omega_i^g =g^{-1}\ast\omega_i\ast g - g^{-1}\ast\partial_i g.
\end{eqnarray}
Then
\[
\begin{aligned}
e\ast \omega_i^g\ast (1-e) &= g^{-1}\ast e\ast \omega_i
\ast(1-e)\ast g - g^{-1}\ast e\ast
\partial_i(g)\ast (1-e).
\end{aligned}
\]
By \eqref{connect-property},
\[
\begin{aligned}
g^{-1}\ast e\ast\omega_i\ast (1-e)\ast g &= - g^{-1}\ast e\ast\partial_i(e)\ast g \\
&= -g^{-1}\ast e\ast\partial_i(e\ast g)  +  g^{-1}\ast e\ast \partial_i g \\
&=-g^{-1}\ast e\ast\partial_i(g)\ast e - e\ast\partial_i e  +
g^{-1}\ast e\ast\partial_i g \\
&=- e\ast\partial_i e  +  g^{-1}\ast e \ast\partial_i(g)\ast(1-e).
\end{aligned}
\]
Therefore,
\[
e\ast\omega_i^g \ast(1-e) = - e\ast\partial_i e.
\]
This shows that the $\omega_i^g$ satisfy the condition
\eqref{connect-property}, thus form a connection on $\cM$.

Now for any given $g\in\cG$,  define the maps $\nabla_i^g$ on $\cM$
by
\[
\nabla_i^g\zeta =\partial_i\zeta + \zeta\ast \omega_i^g, \quad
\forall \zeta.
\]
Also, let $\cR_{i j}^g=\partial_i \omega_j^g - \partial_j \omega_i^g
-[\omega_i^g, \, \omega_j^g]_\ast$ be the curvature corresponding to
the connection $\omega_i^g$. Then we have the following result.
\begin{lemma}\label{gauge} Under a gauge transformation procured by
$g\in\cG$,
\[
\begin{aligned} &\nabla_i^g(\zeta\ast g) =\nabla_i(\zeta)\ast g, \quad
\forall \zeta\in\cM; \\ &\cR_{i j}^g = g^{-1}\ast \cR_{i j} \ast g.
\end{aligned}
\]
\end{lemma}
\begin{proof}
Note that
\[
\nabla_i^g(\zeta\ast g) = \partial_i(\zeta)\ast g + \zeta\ast
\partial_i g + \zeta\ast g\ast \omega_i^g = (\partial_i\zeta +\zeta\ast
\omega_i)\ast g.
\]
This proves the first formula.

To prove the second claim, we use the following formulae
\[
\begin{aligned}
\partial_i \omega_j^g - \partial_j \omega_i^g
&= g^{-1}\ast ( \partial_i \omega_j - \partial_j \omega_i)\ast g -
\partial_i(g^{-1}) \ast\partial_j g + \partial_j(g^{-1}) \ast\partial_i g\\
& + [\partial_i(g^{-1})\ast g,\ g^{-1}\ast\omega_j\ast g]_\ast -
[\partial_j (g^{-1})\ast g,\ g^{-1}\ast\omega_i\ast g]_\ast; \\
[\omega_i^g, \ \omega_j^g]_\ast&= g^{-1}\ast[\omega_i, \
\omega_j]_\ast \ast g -
\partial_i(g^{-1}) \ast\partial_j g + \partial_j(g^{-1})\ast \partial_i g\\
& + [\partial_i(g^{-1})\ast g,\ g^{-1}\ast\omega_j\ast g]_\ast -
[\partial_j (g^{-1})\ast g,\ g^{-1}\ast\omega_i\ast g]_\ast.
\end{aligned}
\]
Combining these formulae together we obtain $\cR_{i j}^g = g^{-1}
\cR_{i j} g$. This completes the proof of the lemma.
\end{proof}

\subsection{Vector bundles associated to right projective modules}

Connections and curvatures can be introduced for the right bundle
$\tilde\cM=e\ast\rAm$ in much the same way. Let
$\tilde\omega_i\in\bM_m(\cA)$ ($i=1, 2, \dots, n$) be matrices
satisfying the condition that
\begin{eqnarray}\label{connection-right}
(1-e)\ast\tilde\omega_i\ast e = \partial_i(e)\ast e.
\end{eqnarray}
Then we can introduce a connection consisting of the right covariant
derivatives $\tilde\nabla_i$ ($i=1, 2, \dots, n$) on $\tilde\cM$
defined by
\[
\begin{aligned}
\tilde\nabla_i: \tilde\cM \longrightarrow \tilde\cM,  &\quad &\xi
\mapsto \tilde\nabla_i\xi = \partial_i\xi - \tilde\omega_i\ast\xi.
\end{aligned}
\]
It is easy to show that $\tilde\nabla_i(\xi \ast a) =
\tilde\nabla_i(\xi)\ast a + \xi\ast \partial_ia$ for all $a\in\cA$.

Note that if $\tilde\omega_i$ is equal to $\partial_i e$ for each
$i$, the condition \eqref{connection-right} is satisfied. We call
them  the {\em canonical connection} on $\tilde\cM$.

Returning to a general connection $\tilde\omega_i$, we define the
associated curvature by
\[
\tilde\cR_{i j} = \partial_i\tilde\omega_j -
\partial_j\tilde\omega_i -[\tilde\omega_i, \ \tilde\omega_j]_\ast.
\]
Then for all $\xi\in\tilde\cM$, we have
\[ [\tilde\nabla_i, \ \tilde\nabla_j]\xi = -\tilde\cR_{i j}\ast\xi. \]
We further define the covariant derivatives of $\tilde\cR_{i j}$ by
\[ \tilde\nabla_k\tilde\cR_{i j} = \partial_k \tilde\cR_{i j} +
\tilde\omega_k \ast\tilde\cR_{i j}- \tilde\cR_{i j}\ast
\tilde\omega_k.
\]
Then we have the following result.
\begin{lemma} The curvature on the right bundle $\tilde\cM$ satisfies the
Bianchi identity
\[\tilde\nabla_i\tilde\cR_{j k} +
\tilde\nabla_j\tilde\cR_{k i}+ \tilde\nabla_k\tilde\cR_{i j}=0. \]
\end{lemma}
By direct calculations we can also prove the following result:
\[
[\tilde\nabla_k, [\tilde\nabla_i, \ \tilde\nabla_j]]\xi
=-\tilde\nabla_k(\tilde\cR_{i j})\ast \xi, \quad \forall
\xi\in\tilde\cM.
\]

Consider the gauge group $\cG$ defined by \eqref{gauge-group}, which
has a right action on $\tilde\cM$:
\[
\tilde\cM \times \cG \longrightarrow \tilde\cM, \quad \xi\times g
\mapsto \xi\cdot g := g^{-1}\ast \xi.
\]
Under a gauge transformation procured by $g\in\cG$,
\[
\tilde\omega_i \mapsto \tilde\omega_i^g := g^{-1} \ast
\tilde\omega_i \ast g +
\partial_i(g^{-1})  \ast g.
\]
The connection $\tilde\nabla_i^g$ on $\tilde\cM$ defined by
\[ \tilde\nabla_i^g\xi = \partial_i\xi - \tilde\omega_i^g \ast\xi \]
satisfies the following relation for all $\xi\in\tilde\cM$:
\[
\tilde\nabla_i^g(g^{-1}\ast\xi) = g^{-1}\ast\tilde\nabla_i\xi.
\]
Furthermore, the gauge transformed curvature
\[
\tilde\cR_{i j}^g := \partial_i \tilde\omega_j^g - \partial_j
\tilde\omega_i^g - [\tilde\omega_i^g, \tilde\omega_j^g]_\ast
\]
is related to $\tilde\cR_{i j}$ by
\[ \tilde\cR_{i j}^g = g^{-1}\ast \tilde\cR_{i j}\ast g.\]

Given any $\Lambda\in\bM_m(\cA)$, we can define the $\cA$-bimodule
map
\begin{eqnarray}\label{form}
\langle \ , \ \rangle:
\cM\otimes_{\R[[\barh]]}\tilde\cM\longrightarrow \cA, \quad
\zeta\otimes\xi \mapsto \langle \zeta , \xi \rangle
=\zeta\ast\Lambda\ast\xi,
\end{eqnarray}
where $\zeta \ast\Lambda \ast\xi$ is defined by matrix
multiplication. We shall say that the bimodule homomorphism is {\em
gauge invariant} if for any element $g$ of the gauge group $\cG$,
\[
\langle \zeta\cdot g, \xi\cdot g \rangle = \langle \zeta, \xi
\rangle, \quad \forall \zeta\in\cM, \ \xi\in\tilde\cM.
\]
Also, the bimodule homomorphism is said to be {\em compatible with
the connections} $\omega_i$ on $\cM$ and $\tilde\omega_i$ on
$\tilde\cM$ if for all $i=1, 2, \dots, n$
\[
\partial_i \langle \zeta, \xi
\rangle = \langle \nabla_i\zeta, \xi \rangle + \langle \zeta,
\tilde\nabla_i\xi \rangle, \quad \forall \zeta\in\cM, \
\xi\in\tilde\cM.
\]

\begin{lemma}\label{compatibility}
Let $\langle \ , \ \rangle:
\cM\otimes_{\R[[\barh]]}\tilde\cM\longrightarrow \cA$ be an
$\cA$-bimodule homomorphism defined by \eqref{form} with a given
$m\times m$-matrix $\Lambda$ with entries in $\cA$. Then
\begin{enumerate}
\item \label{form-gauge} $\langle \ , \ \rangle$ is gauge invariant if
$g\ast\Lambda\ast g^{-1} = \Lambda$ for all $g\in\cG$;
\item \label{form-gauge2} $\langle \ , \ \rangle$ is compatible with the connections
$\omega_i$ on $\cM$ and $\tilde\omega_i$ on $\tilde\cM$ if for all
$i$,
\[
e\ast(\partial_i\Lambda - \omega_i \ast\Lambda +  \Lambda
\ast\tilde\omega_i)\ast e =0.
\]
\end{enumerate}
\end{lemma}
\begin{proof}
Note that $\langle \zeta\cdot g , \xi\cdot g \rangle  = \zeta \ast
g\ast \Lambda\ast g^{-1}\ast \xi$ for any $g\in\cG$, $\zeta\in\cM$
and $\xi\in\tilde\cM$. Therefore $\langle \zeta\cdot g , \xi\cdot g
\rangle = \langle \zeta, \xi\rangle$ if $g \ast\Lambda \ast
g^{-1}=\Lambda$. This proves part (\ref{form-gauge}).

Now
$
\partial_i \langle \zeta, \xi\rangle = \langle \partial_i\zeta,
\xi\rangle + \langle\zeta,  \partial_i\xi\rangle +
\zeta\ast(\partial_i\Lambda - \omega_i\ast\Lambda +
\Lambda\ast\tilde\omega_i)\ast\xi. $
Thus if $\Lambda$ satisfies the condition of part
(\ref{form-gauge2}), then $\langle \ , \ \rangle$ is compatible with
the connections.
\end{proof}
%\end{document}

%
\subsection{Canonical connections and fibre metric}\label{canonical}

Let us consider in detail the canonical connections on $\cM$ and
$\tilde\cM$ given by
\[ \omega_i = - \partial_i e, \quad \tilde\omega_i=\partial_i e. \]
A particularly nice feature in this case is that the corresponding
curvatures on the left and right bundles coincide. We have the
following formula:
\[ \cR_{i j} = \tilde\cR_{i j}= - [\partial_i e, \ \partial_j e]_\ast.\]

Now we consider a special case of the $\cA$-bimodule map defined by
equation \eqref{form}.
\begin{definition}\label{fibre-metric}
Denote by $\g: \cM\otimes_{\R[[\barh]]} \tilde\cM \longrightarrow
\cA$ the map defined by \eqref{form} with $\Lambda$ being the
identity matrix. We shall call $\g$ the {\em fibre metric} on $\cM$.
\end{definition}
\begin{lemma}\label{metric}
The fibre metric $\g$ is gauge invariant and is compatible with the
standard connections.
\end{lemma}
\begin{proof}
Since $\Lambda$ is the identity matrix in the present case, it
immediately follows from Lemma \ref{compatibility}
(\ref{form-gauge}) that $\g$ is gauge invariant. Note that
$e\ast\partial_i(e)\ast e =0$ for all $i$. Using this fact in Lemma
\ref{compatibility} (\ref{form-gauge2}), we easily see that $\g$ is
compatible with the standard connections.
\end{proof}

\section{Embedded noncommutative spaces}\label{surfaces}
In this section we study explicit examples of idempotents and
related projective modules. They correspond to the noncommutative
spaces introduced in \cite{CTZZ}. The main result here is a
reformulation of the theory of embedded noncommutative spaces
\cite{CTZZ} in the framework of Section \ref{bundles} in terms of
projective modules.

\subsection{Embedded noncommutative spaces}

We shall consider  only embedded spaces with Euclidean signature.
The Minkowski case is similarly, which we shall briefly allude to in
Remark \ref{Mink} at the end of this section. Given
$X=\begin{pmatrix} X^1 & X^2 &\dots & X^m
\end{pmatrix}$ in $\lAm$, we define an ($n\times n$)-matrix $(g_{i
j})_{i, j=1, 2, \dots, n}$ with entries given by
\[
g_{i j}=\sum_{\alpha=1}^m \partial_i X^\alpha \ast \partial_j
X^\alpha.
\]
Following \cite{CTZZ}, we shall call $X$ a {\em noncommutative
space} embedded in $\cA^m$ if the matrix $(g_{i j})$ is invertible.

For a given noncommutative space $X$, we denote by $(g^{i j})$ the
inverse matrix of $(g_{i j})$ with $g_{i j}\ast g^{j k} = g^{k
j}\ast g_{j i} = \delta^k_i$ for all $i$ and $k$. Here Einstein's
summation convention is used, and we shall continue to use this
convention throughout the paper. Let
\[
E_i=\partial_i X,  \quad \tilde E^i=(E_j)^t\ast g^{j i}, \quad
E^i=g^{i j}\ast E_j,
\]
for $i=1, 2, \dots, n$, where $(E_i)^t= \begin{pmatrix}\partial_i X^1 \\
\partial_i X^2
\\ \vdots \\ \partial_i X^m\end{pmatrix}$ denotes the transpose of $E_i$. Define $e\in\bM_m(\cA)$ by
\begin{eqnarray}\label{idempotent}
\begin{aligned}
e:&= \tilde{E}^j\ast E_j \\
&=\begin{pmatrix}
\partial_i X^1\ast g^{i j}\ast \partial_j X^1  & \partial_i X^1\ast g^{i j}\ast \partial_j X^2
& \dots &\partial_i X^1\ast g^{i j}\ast \partial_j X^m \\
\partial_i X^2 \ast g^{i j}\ast  \partial_j X^1  & \partial_i X^2  \ast g^{i j}\ast  \partial_j X^2
& \dots &\partial_i X^2  \ast g^{i j}\ast  \partial_j X^m \\
\dots & \dots & \dots & \dots \\
\partial_i X^m  \ast g^{i j}\ast  \partial_j X^1  & \partial_i X^m  \ast g^{i j}\ast  \partial_j X^2
& \dots &\partial_i X^m  \ast g^{i j}\ast  \partial_j X^m
\end{pmatrix}.
\end{aligned}
\end{eqnarray}
We have the following results.
\begin{proposition}
\begin{enumerate}
\item  Under matrix multiplication, $E_i \ast \tilde{E}^j =\delta^j_i$ for all $i$ and $j$.
\item The $m\times m$ matrix $e$ satisfies $e \ast  e=e$, that is, it is an
idempotent in $\bM_m(\cA)$.
\item The left and right projective $\cA$-modules $\cM=\lAm \ast  e$
and $\tilde \cM=e \ast \rAm$ are respectively spanned by $E_i$ and
$\tilde{E}^i$. More precisely, we have
\[
\cM=\{a^i \ast  E_i \mid a^i\in\cA\}, \quad  \tilde\cM=\{\tilde{E}^i
 \ast  b_i \mid b_i \in\cA\}.
\]
\end{enumerate}
\end{proposition}
\begin{proof}
Note that $g_{i j} = E_i \ast  (E_j)^t$. Thus $E_i \ast  \tilde E^j
= E_i \ast  (E_k)^t \ast  g^{k j} = \delta_i^j$. It then immediately
follows that
\[
e \ast e = \tilde{E^i} \ast  \left(E_i  \ast \tilde{E^j}\right) \ast
E_j = \tilde{E^i}  \ast \delta^j_i  \ast E_j = e.
\]
Obviously $\cM\subset\{a^i \ast  E_i \mid a^i\in\cA\}$ and $
\tilde\cM\subset\{\tilde{E}^i \ast  b_i\mid b_i \in\cA\}$. By the
first part of the proposition, we have
\[
\begin{aligned}
a^i \ast  E_i \ast  e = a^i \ast  \left(E_i \ast  \tilde{E}^j\right)
 \ast E_j = a^j \ast  E_j,
\\
e \ast \tilde{E}^j \ast  b_j= \tilde{E}^i
 \ast \left(E_i \ast \tilde{E}^j\right) \ast  b_j = \tilde{E}^i \ast  b_i.
\end{aligned}
\]
This proves the last claim of the proposition.
\end{proof}
It is also useful to observe that $\tilde\cM=\{(E_i)^t \ast  b_i\mid
b_i \in\cA\}$ since $(g_{i j})$ is invertible.

We shall denote $\cM$ and $\tilde\cM$ respectively by $TX$ and
$\tilde TX$, and refer to them as the {\em left} and {\em right
tangent bundles} of the noncommutative space $X$. Note that the
definition of the tangent bundles coincides with that in
\cite{CTZZ}.

\begin{definition}
Call the fibre metric  $\g: TX\otimes_{\R[[\barh]]} \tilde TX
\longrightarrow \cA$ defined in Definition \ref{fibre-metric} the
{\em metric} of the noncommutative space $X$.
\end{definition}

The proposition below in particular shows that $\g$ agrees with the
metric of the embedded noncommutative space defined in \cite{CTZZ}
in a geometric setting.

\begin{proposition}
For any $\zeta=a^i \ast  E_i \in TX$ and $\xi= (E_j)^t \ast b^j\in
\tilde TX$ with $a_i, b_j\in \cA$,
\[
\g: \zeta \otimes \xi \mapsto \g(\zeta, \xi)=a^i  \ast  g_{i j} \ast
b^j.
\]
In particular, $\g(E_i, (E_j)^t) = g_{i j}$.
\end{proposition}
\begin{proof}
Recall from Definition \ref{fibre-metric} that $\g$ is defined by
\eqref{form} with $\Lambda$ being the identity matrix. Thus for any
$\zeta=a^i \ast  E_i \in TX$ and $\xi= (E_j)^t \ast b^j\in \tilde
TX$ with $a_i, b_j\in \cA$,
\[\g(\zeta, \xi)= a^i \ast  E_i  \ast (E_j)^t  \ast b^j = a^i  \ast g_{i j} \ast b^j. \]
This completes the proof.
\end{proof}

Let us now equip the left and right tangent bundles with the {\em
canonical connections} given by $\omega_i= -\tilde\omega_i
=-\partial_i e$, and denote the corresponding covariant derivatives
by
\[
\nabla_i: T X\longrightarrow T X, \quad \tilde\nabla_i: \tilde T
X\longrightarrow \tilde T X.
\]
In principle, one can take arbitrary connections for the tangent
bundles, but we shall not allow this option in this paper.

The following elements of $\cA$ are defined in \cite{CTZZ},
\begin{eqnarray*}
_c\Gamma_{i j l} = \frac{1}{2} \left(\partial_i g_{j l} +
\partial_j g_{l i}
-\partial_l g_{j i} \right), &\quad& \Upsilon_{i j l} = \frac{1}{2}
\left(\partial_i(E_j) \ast  (E_l)^t - E_l \ast \partial_i (E_j)^t\right),\\
\Gamma_{i j l} =  {}_c\Gamma_{i j l} + \Upsilon_{i j
 l}, &\quad& \tilde\Gamma_{i j l} =
{}_c\Gamma_{i j l}- \Upsilon_{i j l},
\end{eqnarray*}
where $\Upsilon_{i j k}$ was referred to as the noncommutative
torsion.  Set \cite{CTZZ}
\begin{eqnarray}\label{Christoffel}
\Gamma_{i j}^k =  \Gamma_{i j l} \ast  g^{l k}, \quad
\tilde\Gamma_{i j}^k = g^{k l} \ast  \tilde\Gamma_{i j l}.
\end{eqnarray}
Then we have the following result.
\begin{lemma}\label{Gamma}
\begin{eqnarray}
\nabla_i E_j= \Gamma_{i j}^k  \ast  E_k, && \tilde \nabla_i \tilde
E^j= - \tilde E^k\ast \Gamma_{k i}^j.
\end{eqnarray}
\end{lemma}
\begin{proof}
Consider the first formula. Write $\partial_i e =
\partial_i(\tilde{E}^k) \ast E_k + \tilde{E}^k \ast \partial_iE_k$. We have
\[
\begin{aligned}
\nabla_i E_j &= \partial_i E_j - E_j \partial_i \ast  e \\
&= \partial_i E_j - \left(\partial_i(E_j  \ast e) - \partial_i(E_j)
 \ast e
\right) \\
&=  \partial_i(E_j) \ast \tilde{E}^k  \ast E_k.
\end{aligned}
\]
It was shown in \cite{CTZZ} that $\Gamma_{i j}^k =\partial_i(E_j)
\ast \tilde{E}^k$. This immediately leads to the first formula. The
proof for the second formula is essentially the same.
\end{proof}

Note that Lemma \ref{Gamma} can be re-stated as
\[
\nabla_i E^j = - \tilde\Gamma_{i k}^j\ast E^k, \quad \tilde \nabla_i
(E_j)^t= (E_k)^t \ast \tilde \Gamma_{i j}^k.
\]

By using Lemma \ref{metric} and Lemma \ref{Gamma}, we can easily
prove the following result, which is equivalent to \cite[Proposition
2.7]{CTZZ}.
\begin{proposition}\label{prop:compatible}
The connections are metric compatible in the sense that
\begin{eqnarray}\label{compatible}
\partial_i \g(\zeta, \xi)=\g(\nabla_i\zeta, \xi) + \g(\zeta,
\tilde\nabla_i\xi), &\quad& \forall \zeta\in TX, \ \xi\in \tilde TX.
\end{eqnarray}
\end{proposition}
For $\zeta=E_j$ and $\xi=(E_k)^t$, we obtain from \eqref{compatible}
the following result for all $i, j, k$:
\begin{eqnarray}\label{compatible1}
\partial_i g_{j k} -\Gamma_{i j k} - \tilde\Gamma_{i k j} =0.
\end{eqnarray}
This formula is in fact equivalent to Proposition
\ref{prop:compatible}.

Define
\begin{eqnarray}\label{R-tensor}
R^l_{k i j} = E_k \ast \cR_{i j} \ast \tilde E^l, \quad \tilde
R^l_{k i j} =- g^{l q} \ast  E_q \ast \cR_{i j} \ast \tilde E^p \ast
g_{p k}.
\end{eqnarray}
Using $\cR_{i j} = \tilde\cR_{i j} = - [\partial_i e,
\partial_j e]_\ast $, we can show by some lengthy calculations that
\begin{eqnarray}\label{curvature1}
\begin{aligned} R_{k i j}^l &=& -\partial_j\Gamma_{i k}^l -
\Gamma_{i k}^p  \ast  \Gamma_{j p}^l + \partial_i\Gamma_{j k}^l
+\Gamma_{j k}^p  \ast \Gamma_{i p}^l ,\\
\tilde R_{k i j}^l &=& -\partial_j\tilde\Gamma_{i k}^l -
\tilde\Gamma_{j p}^l \ast \tilde\Gamma_{i k}^p
+\partial_i\tilde\Gamma_{j k}^l + \tilde\Gamma_{i p}^l
 \ast \tilde\Gamma_{j k}^p,
\end{aligned}
\end{eqnarray}
which are the {\em Riemannian curvatures} of the left and right
tangent bundles of the noncommutative space $X$ given in \cite[Lemma
2.12 and \S 4]{CTZZ}. Therefore,
\begin{eqnarray}\label{curvature}
{[\nabla_i, \nabla_j]} E_k = R_{k i j}^l  \ast E_l, & \quad &
{[}\tilde\nabla_i, \tilde\nabla_j{]}(E_k)^t = (E_l)^t  \ast \tilde
R_{k i j}^l,
\end{eqnarray}
recovering the relations \cite[(2.13)]{CTZZ} and their
generalisations \cite[\S 4]{CTZZ} to arbitrary $m\ge n$.

\begin{remark}\label{Mink}
We comment briefly on noncommutative spaces with  Minkowski
signatures embedded in higher dimensions \cite{CTZZ}. Let
$\eta=diag(-1, \dots, -1, 1, \dots, 1)$ be a diagonal $m\times m$
matrix with $p$ of the diagonal entries being $-1$, and $q=m-p$ of
them being $1$. Given $X=\begin{pmatrix} X^1 & X^2 &\dots & X^m
\end{pmatrix}$ in $\lAm$, we define an $n\times n$ matrix $(g_{i
j})_{i, j=1, 2, \dots, n}$ with entries
\[
g_{i j}=\sum_{\alpha=1}^m \partial_i X^\alpha  \ast \eta_{\alpha
\beta}  \ast \partial_j X^\beta.
\]
We call $X$ a {\em noncommutative space} embedded in $\cA^m$ if the
matrix $(g_{i j})$ is invertible. Denote its inverse matrix by
$(g^{i j})$. Now the idempotent which gives rise to the left and
right tangent bundles of $X$ is given by
\[
e=  \eta (E_i)^t  \ast g^{i j} \ast  E_j,
\]
which obviously satisfies $E_i \ast  e = E_i$ for all $i$.  The
fibre metric of Definition \ref{metric} yields a metric on the
embedded noncommutative surface $X$.
\end{remark}

\subsection{Example}\label{example} %%

We analyze an embedded noncommutative surface of Euclidean signature
arising from the quantisation of a time slice of the Schwarzschild
spacetime. While the main purpose here is to illustrate how the
general theory developed in previous sections works, the example is
interesting in its own right.

Let us first specify the notation to be used in this section. Let
$t^1=r$, \ $t^2=\theta$ and $t^3=\phi$, with $r>2m$, $\theta\in (0,
\pi)$, and $\phi\in (0, 2\pi)$. We deform the algebra of functions
in these variables by imposing the Moyal product defined by
\eqref{multiplication} with the following anti-symmetric matrix
\begin{eqnarray*}\label{asy-matrix}
\left(\theta _{i j}\right)_{i, j=1}^3=\left(\begin{array}{cccc}
    0&  0&  0\\
    0&  0&  1\\
    0&  -1&  0
\end{array}\right).
\end{eqnarray*}
Note that the functions depending  only on the variable $r$ are
central in the Moyal algebra $\cA$. We shall write the usual
pointwise product of two functions $f$ and $g$ as $f g$, but write
their Moyal product as $f\ast g$.

Consider $X=\begin{pmatrix} X^1 & X^2 & X^3 & X^4\end{pmatrix}$ given by
\begin{eqnarray}
\begin{aligned}
X^1=f(r) \quad \text{with} \quad (f')^2 +1=\left(1-\frac{2m}{r}\right)^{-1},\\
X^2=r \sin\theta \cos\phi,\quad X^3=r \sin\theta \sin\phi, \quad
X^4=r \cos\theta.
\end{aligned}
\end{eqnarray}
Simple calculations yield
\[
\begin{aligned}
E_1&=\partial_r X=\begin{pmatrix}f' &\sin\theta \cos\phi &\sin\theta \sin\phi &\cos\theta\end{pmatrix}, \\
E_2&=\partial_\theta X=\begin{pmatrix}0 &r\cos\theta \cos\phi
&r\cos\theta \sin\phi
&-r\sin\theta\end{pmatrix},\\
E_3&=\partial_\phi X=\begin{pmatrix}0 &-r\sin\theta \sin\phi
&r\sin\theta \cos\phi &0\end{pmatrix}.
\end{aligned}
\]
Using these formulae, we obtain the following expressions for the
components of the metric of the noncommutative surface $X$:
\begin{eqnarray}\label{deformed-Sch}
\begin{aligned}
g_{1 1}=&\left(1-\frac{2m}{r}\right)^{-1}
\left[1-\left(1-\frac{2m}{r}\right)
          \cos(2\theta)\sinh^2\barh\right],\\
g_{1 2} =    &g_{2 1} = r\sin(2\theta)\sinh^2\barh,\\
g_{2 2} =    &r^2
    \left[1+\cos(2\theta)\sinh^2\barh\right],\\
g_{2 3} = &-g_{3 2}
=-r^2\cos(2\theta)\sinh\barh\cosh\barh,\\
g_{1 3} = &-g_{3 1}
=-r\sin(2\theta)\sinh\barh\cosh\barh,\\
g_{3 3} =&r^2
    \left[\sin^2\theta - \cos(2\theta)\sinh^2\barh\right].
\end{aligned}
\end{eqnarray}
In the limit $\barh\rightarrow 0$, we recover the spatial components
of the Schwarzschild metric.  Observe that the noncommutative
surface still reflects the characteristics of the Schwarzschild
spacetime in that there is a time slice of the Schwarzschild black
hole with the event horizon at $r = 2m$.

Since the metric $(g_{i j})$ depends on $\theta$ and $r$ only, and
the two variables commute, the inverse $(g^{i j})$ of the metric can
be calculated in the usual way as in the commutative case. Now the
components of the idempotent $e=(e_{i j})= (E_i)^t * g^{i j} * E_j$
are given by the following formulae:
\begin{eqnarray*}
\begin{aligned}
e_{11} =&\frac{2m}{r} + \frac{2m(2m - r)(2 + \cos 2\theta)}{r^2} \barh^2+ O(\barh^3),\\
e_{12} =&\frac{m \cos\phi \sin\theta}{r \sqrt{\frac{m}{-4m + 2r}}}
        - \frac{2m\cos\theta\sin\phi }{r\sqrt{\frac{m}{-4m + 2r}}}
        \barh\\
        &+
  \frac{m( 4m + r + 2m\cos 2\theta ) \cos\phi \sin\theta }{r^2 \sqrt{\frac{m}{-4m + 2r}}} \barh^2+ O(\barh^3)\\
e_{13}=&\frac{m\sin\theta\sin\phi}{r\sqrt{\frac{m}{-4m + 2r}}} +
\frac{2m\cos\theta\cos\phi }{r{\sqrt{\frac{m}{-4m + 2r}}}} \barh \\
&+ \frac{m( 4m + r + 2m\cos 2\theta) \sin\theta \sin\phi }{r^2 {\sqrt{\frac{m}{-4m + 2r}}}} \barh^2 + O(\barh^3)\\
e_{14}=&\frac{m\cos\theta}{r{\sqrt{\frac{m}{-4m + 2r}}}} +
\frac{m\cos\theta( 4m - r + 2m\cos 2\theta)}
   {r^2 {\sqrt{\frac{m}{-4m + 2r}}}} \barh^2 + O(\barh^3)\\
%\end{aligned}
%\end{eqnarray*}
%\begin{eqnarray*}
%\begin{aligned}
e_{21}=&\frac{m\cos\phi\sin\theta}{r{\sqrt{\frac{m}{-4m + 2r}}}} +
\frac{2m\cos\theta \sin\phi }{r{\sqrt{\frac{m}{-4\,m +
2\,r}}}}\barh\\ &+\frac{m ( 4m + r + 2m\cos 2\theta) \cos\phi
\sin\theta }{r^2 {\sqrt{\frac{m}{-4m + 2r}}}} \barh^2
+ O(\barh^3)\\
e_{22}=&1-\frac{2 m\sin^2 \theta \cos^2 \phi }{r} \\
&+ \frac{m}{2 r^2}\Big[2r + 2m\cos 4\theta \cos^2 \phi - 6 m\cos^2
\phi \\
&+2\cos 2\theta (m + 8r + ( m - r)\cos 2\phi ) \Big] \barh^2 + O(\barh^3)\\
e_{23}=&-\frac{m \sin^2 \theta \sin 2\phi}{r} - \frac{3m\sin 2\theta }{r} \barh\\
&+ \frac{m( 2( m - r)\cos 2\theta + m( -3 + \cos 4\theta ) ) \sin
2\phi}{2r^2} \barh^2+
  O(\barh^3)\\
e_{24}=&\frac{-2m \cos\theta \cos\phi \sin\theta}{r} - \frac{m( 1 + 3\cos 2\theta)\sin\phi }{r} \barh\\
&-\frac{m ( 8 m + 5r + 4m \cos 2\theta ) \cos\phi \sin 2\theta }{2 r^2} \barh^2+ O(\barh^3)\\
\end{aligned}
\end{eqnarray*}
\begin{eqnarray*}
\begin{aligned}
e_{31}=&\frac{m\sin\theta\sin\phi}{r{\sqrt{\frac{m}{-4m + 2r}}}}
-\frac{2m\cos\theta\cos\phi }{r{\sqrt{\frac{m}{-4m + 2r}}}} \barh\\
&+   \frac{m( 4m + r + 2m\cos 2\theta) \sin\theta \sin\phi }{r^2 {\sqrt{\frac{m}{-4m + 2r}}}} \barh^2 + O(\barh^3)\\
e_{32}=&-\frac{m \sin^2 \theta \sin 2\phi}{r} + \frac{3m\sin 2\theta }{r} \barh \\
&+
  \frac{m( 2( m - r)\cos 2\theta + m( -3 + \cos 4\theta ) ) \sin 2\phi }{2r^2} \barh^2+
  O(\barh^3)\\
e_{33}=&1-\frac{2 m\sin^2 \theta \sin^2 \phi }{r} \\
&+ \frac{m}{2 r^2}\Big[ 2r + 2m\cos 4\theta \sin^2 \phi - 6m\sin^2
\phi \\
&+ 2\cos 2\theta (m + 8r - ( m - r)\cos 2\phi ) \Big] \barh^2 + O(\barh^3)\\
e_{34}=&\frac{-2m\cos\theta \sin\theta \sin\phi }{r} + \frac{m( 1 + 3\cos 2\theta ) \cos\phi }{r} \barh \\
&-
  \frac{m( 8m + 5r + 4m \cos 2\theta) \sin 2\theta \sin\phi }{2 r^2} \barh^2+ O(\barh^3)\\
\end{aligned}
\end{eqnarray*}
\begin{eqnarray*}
\begin{aligned}
e_{41}=&\frac{m\cos\theta}{r{\sqrt{\frac{m}{-4m + 2r}}}} +
\frac{m\cos\theta( 4m - r + 2m\cos 2\theta)}
   {r^2 {\sqrt{\frac{m}{-4m + 2r}}}} \barh^2 + O(\barh^3)\\
e_{42}=&\frac{-2m \cos\theta \cos\phi \sin\theta}{r} + \frac{m( 1 + 3\cos 2\theta)\sin\phi }{r} \barh\\
&-
  \frac{m ( 8 m + 5r + 4m \cos 2\theta ) \cos\phi \sin 2\theta }{2 r^2} \barh^2+ O(\barh^3)\\
e_{43}=&\frac{-2m\cos\theta \sin\theta \sin\phi }{r} -\frac{m( 1 + 3\cos 2\theta ) \cos\phi }{r} \barh \\
&-
  \frac{m( 8m + 5r + 4m \cos 2\theta) \sin 2\theta \sin\phi }{2 r^2} \barh^2 + O(\barh^3)\\
e_{44}=&1-\frac{2m \cos^2\theta}{r} + \frac{4m \cos^2 \theta ( -2m +
r - m\cos 2\theta )}{r^2} \barh^2 +O(\barh^3)
\end{aligned}
\end{eqnarray*}

Let us write $ e = e_0 + \barh e_1 + \barh^2 e_2 +\cdots$.  Then
inspecting the formulae we see that the matrices $e_0$ and $e_2$ are
symmetric, while $e_1$ is skew symmetric. This is no coincidence;
rather it is a consequence of properties of $X$ under the bar
involution, which will be discussed in Section \ref{bar}.

Here we refrain from presenting the result of the Mathematica
computation for the curvature $\cR_{i j}=-[\partial_i e,
\partial_j e]$, which is very complicated and
not terribly illuminating. However, we mention that in \cite{WZZ}
a quantisation of the Schwarzschild spacetime was carried out
(for a particular choice of $\Theta$),
and the resulting noncommutative differential geometry
was studied in detail. In particular, the
metric, Christoffel symbols, Riemannian and Ricci curvatures were
explicitly worked out. We refer to that paper for details.

\section{General coordinate transformations}\label{transformations}

We now  return to the general setting of Section \ref{bundles} to
investigate ``general coordinate transformations".
Our treatment follows closely \cite[\S V]{CTZZ} and makes
use of general ideas of \cite{Ge, D, Ko}.
We should point out that the material presented is part of
an attempt of ours to develop
a notion of ``general covariance" in the noncommutative setting. This is
an important matter which deserves a thorough investigation. We hope that
the work presented here will prompt further studies.

Let $(\cA, \mu)$ be a Moyal algebra of smooth functions on the open
region $U$ of $\R^n$ with coordinate $t$. This algebra is defined
with respect to a constant skew symmetric matrix $\theta=(\theta_{i
j})$. Let $\Phi: U\longrightarrow U$ be a diffeomorphism of $U$ in
the classical sense. We denote \[ u^i=\Phi^i(t), \] and refer to
this as a {\em general coordinate transformation} of $U$.

Denote by $\cA_u$ the sets of smooth functions of $u=(u^1, u^2,
\dots, u^n)$. The map $\Phi$ induces an $\R[[\barh]]$-module
isomorphism $\phi=\Phi^*: \cA_u\longrightarrow \cA$ defined for any
function $f\in\cA_u$ by
\[\phi(f)(t) =f(\Phi(t)).\]
We define the $\R[[\barh]]$-bilinear map
\[
\mu_u: \cA_u\otimes \cA_u \longrightarrow \cA_u, \quad \mu_u(f,
g)= \phi^{-1} \mu_t(\phi(f), \phi(g)).
\]
Then it is well-known \cite{Ge} that $\mu_u$ is associative.
Therefore, we have the associative algebra isomorphism
\[
\phi: (\cA_u, \mu_u) \stackrel{\sim}{\longrightarrow} (\cA_t,
\mu_t).
\]
We say that the two associative algebras are {\em gauge equivalent}
by adopting the terminology of \cite{D}.

Following \cite{CTZZ}, we define $\R[[\barh]]$-linear operators
\begin{eqnarray}\label{dphi}
\partial_i^\phi := \phi^{-1}\circ \partial_i\circ \phi: \cA_u \longrightarrow
\cA_u,
\end{eqnarray}
which have the following properties \cite[Lemma 5.5]{CTZZ}:
\[
\begin{aligned}
&\partial_i^\phi \circ \partial_j^\phi - \partial_j^\phi \circ
\partial_i^\phi=0, \\
&\partial_i^\phi\mu_u(f, g) = \mu_u(\partial_i^\phi(f),  g) +
\mu_u(f, \partial_i^\phi(g)), \quad \forall f, g\in \cA_u,
\end{aligned}
\]
where the second relation is the Leibniz rule for $\partial_i^\phi$.
Recall that this Leibniz rule played a crucial role in the
construction of noncommutative spaces over $(\cA_u, \mu_u)$ in
\cite{CTZZ}.

We shall denote by $\bM_m(\cA_u)$ the set of $m\times m$-matrices
with entries in $\cA_u$. The product of two such matrices will be
defined with respect to the multiplication $\mu_u$ of the algebra
$(\cA_u, \mu_u)$. Then $\phi^{-1}$ acting component wise gives rise
to an algebra isomorphism from $\bM_m(\cA)$ to $\bM_m(\cA_u)$, where
matrix multiplication in $\bM_m(\cA)$ is defined with respect to
$\mu$.

Since we need to deal with two different algebras $(\cA, \mu)$ and
$(\cA_u, \mu_u)$ simultaneously in this section, we write $\mu$ and
the matrix multiplication defined with respect to it by $\ast$ as
before, and use $\ast_u$ to denote $\mu_u$ and the matrix
multiplication defined with respect to it.

Let $e\in\bM_m(\cA)$ be an idempotent. There exists the
corresponding finitely generated projective left (resp. right)
$\cA$-module $\cM$ (resp. $\tilde\cM$). Now $e_u:=\phi^{-1}(e)$ is
an idempotent in $\bM_m(\cA_u)$, that is,
$\phi^{-1}(e)\ast_u\phi^{-1}(e)=\phi^{-1}(e)$. Write
$e_u=(\cE_\alpha^\beta)_{\alpha, \beta=1, \dots, m}$. This
idempotent gives rises to the left projective $\cA_u$-module $\cM_u$
and right projective $\cA_u$-module $\tilde\cM_u$, respectively
defined by
\begin{eqnarray*}
&\cM_u = \left\{\left.\begin{pmatrix}a^\alpha \ast_u\cE_\alpha^1&
a^\alpha \ast_u\cE_\alpha^2 & \dots & a^\alpha \ast_u\cE_\alpha^m
\end{pmatrix}\right| a^\alpha\in \cA_u\right\},\\
&\tilde\cM_u = \left\{\left.\begin{pmatrix}
\cE_1^\beta \ast_u b_\beta \\ \cE_2^\beta\ast_u b_\beta\\
\vdots \\ \cE_m^\beta \ast_u b_\beta
\end{pmatrix}\right| b^\beta\in \cA_u\right\},
\end{eqnarray*}
where $a^\alpha \ast_u \cE_\alpha^\beta = \sum_{\alpha}
\mu_u(a^\alpha, \cE_\alpha^\beta)$ and $\cE_\alpha^\beta\ast_u
b_\beta =\sum_{\beta} \mu_u(\cE_\alpha^\beta, b_\beta)$. Below we
consider the left projective module only, as the right projective
module may be treated similarly.

Assume that we have the left connection
\[
\nabla_i: \cM\longrightarrow \cM, \quad \nabla_i\zeta
=\frac{\partial\zeta}{\partial t^i} + \zeta\ast\omega_i.
\]
Let $\omega_i^u:=\phi^{-1}(\omega_i)$. We have the following result.
\begin{theorem} \label{covariance}
\begin{enumerate}
\item The matrices $\omega_i^u$ satisfy the following relations in
$\bM_m(\cA_u)$:
\[ e_u\ast_u \omega_i^u \ast_u (1- e_u) = - e_u \ast_u  \partial_i^\phi e_u.\]
\item The operators $\nabla^\phi_i$  ($i=1, 2, \dots, n$) defined for all $\eta\in\cM_u$ by
\[ \nabla^\phi_i\eta =\partial_i^\phi\eta + \eta\ast_u \omega_i^u\]
give rise to a connection on $\cM_u$.
\item The curvature of the connection $\nabla^\phi_i$ is given by
\[\cR_{i j}^u = \partial_i^\phi\omega_j^u- \partial_j^\phi\omega_i^u
-\omega_i^u\ast_u \omega_j^u+ \omega_j^u\ast_u \omega_i^u,\] which
is related to the curvature $\cR_{i j}$ of $\cM$ by
\[\cR_{i j}^u = \phi^{-1}(\cR_{i j}).\]
\end{enumerate}
\end{theorem}
\begin{proof}Note that $e_u \ast_u\omega_i^u \ast_u(1- e_u)
=\phi^{-1}(e\ast\omega_i\ast(1-e))$.
We also have $\partial_i^\phi e_u= \phi^{-1}(\frac{\partial
e}{\partial t^i})$, which leads to $e_u\ast_u \partial_i^\phi
e_u=\phi^{-1}(e\ast \phi(\partial_i^\phi e_u))=
\phi^{-1}(e\ast\partial_i e)$. This proves part (1). Part (2)
follows from part (1) and the Leibniz rule for $\partial_i^\phi$.
Straightforward calculations show that the curvature of the
connection $\nabla^\phi_i$ is given by $\cR_{i j}^u
=\partial_i^\phi\omega_j^u- \partial_j^\phi\omega_i^u
-\omega_i^u\ast_u\omega_j^u +  \omega_j^u\ast_u\omega_i^u$. Now
$\partial_i^\phi\omega_j^u = \phi^{-1}(\frac{\partial
\omega_j}{\partial t^i})$, and
$\omega_i^u\ast_u\omega_j^u-\omega_j^u\ast_u\omega_i^u=\phi^{-1}(\omega_i\ast\omega_j)
-\phi^{-1}(\omega_j\ast\omega_i)$. Hence $\cR_{i j}^u =
\phi^{-1}(\cR_{i j})$.
\end{proof}

\begin{remark}\label{classical-rem}
One can recover the usual transformation rules of tensors
under the diffeomorphism group from the commutative limit of Theorem  \ref{covariance}
in a way similar to that in \cite[\S 5.C]{CTZZ}.
\end{remark}

\section{Bar involution and generalised Hermitian structure}\label{bar}

In this section, we study a Moyal algebra analogue of the bar map of
quantum groups, and investigate its implications on noncommutative
geometry. Note that the ring $\R[[\barh]]$ admits an involution that
maps an arbitrary power series $a=\sum_i a_i \barh^i$ in
$\R[[\barh]]$ to $\bar a=\sum_i (-1)^i a_i \barh^i$. We shall call
$\bar a$ the {\em conjugate} of $a$. Note that $\bar a a$ contains
only even powers of $\barh$. We can extend this map to a conjugate
linear anti-involution on the Moyal algebra $\cA$.
\begin{lemma}\label{bar-involution}
Let $\bar{\ }: \cA \longrightarrow \cA$ be the map defined for any
$f=\sum_i f_i \barh^i\in \cA$, where $f_i$ are real functions on
$U$, by $\bar{f}= \sum_i (-1)^i f_i \barh^i$. Then for all $f,
g\in\cA$,
\[ \overline{f\ast g} = \bar{g}\ast \bar{f}.
\]
\end{lemma}
We refer to the map as the {\em bar involution} of the Moyal
algebra. It is an analogue of the well known bar map, sending
$q=\exp(\barh)$ to $q^{-1}$, in the theory of quantum groups, which
plays an important role in the study of canonical (crystal) bases.

The lemma can be easily proven by inspecting \eqref{multiplication}.
Given any rectangular matrix $A=(a_{r s})$ with entries in $\cA$, we
let $A^\dagger$ be the matrix obtained from $A$ by first taking its
transpose then sending every matrix elements to its conjugate. For
example, $\begin{pmatrix}a_1 & b_1 & c_1\\
a_2 & b_2 & c_2\end{pmatrix}^\dagger$ $=$  $\begin{pmatrix}\overline{a_1} & \overline{a_2}\\
\overline{b_1} & \overline{b_2}\\ \overline{c_1} &
\overline{c_2}\end{pmatrix} $. It is clear that if the product
$A\ast B$ of two matrices are defined, then $(A \ast B)^\dagger =
B^\dagger \ast A^\dagger$.

Let $\cA^m=\lAm$ be the $\R[[\barh]]$-module consisting of rows
matrices of length $m$ with entries in $\cA$. We define the form
\begin{eqnarray}\label{form-1}
( \ , \ ): \cA^m\times\cA^m\longrightarrow \cA, \quad \zeta \times
\xi \mapsto (\zeta, \xi):=\zeta\ast\xi^\dagger.
\end{eqnarray}
\begin{lemma}\label{gauge-2}
%The form defined by \eqref{form-1} has the following properties.
\begin{enumerate}
\item For all $\zeta,
\xi\in\cM$ and $a , b\in\cA$,
\[ (\zeta, \xi) =  \overline{(\xi, \zeta)},
\quad (a\ast\zeta, b\ast\xi)=a\ast(\zeta, \xi)\ast\bar{b}.\] Thus in
this sense the form \eqref{form-1} is sesquilinear.
\item $(\zeta, \zeta)=0$ if and only if   $\zeta=0$.
\item For all $\zeta,
\xi\in\cM$ and $A\in\bM_m(\cA)$, we have
\[(\zeta\ast A, \xi) =(\zeta, \xi\ast A^\dagger).\]
\item Let the {\em bar-unitary group} $U_m(\cA)$ over $\cA$ be the
subgroup of $GL_m(\cA)$ defined by $U_m(\cA)=\{g\in GL_m(\cA) \mid
g^\dagger=g^{-1}\}$. Then the form \eqref{form-1} is invariant under
$U_m(\cA)$ in the sense that for all $g\in U_m(\cA)$ and $\zeta,
\xi\in\cM$, \[ (\zeta\ast g, \xi\ast g) = (\zeta, \xi).\]
\end{enumerate}
\end{lemma}
It is straightforward to prove the lemma. Note that part (2) of the
lemma makes the form \eqref{form-1} as nice as a positive definite
hermitian form in the commutative case.

We shall call an idempotent $e\in\bM_m(\cA)$ {\em self-adjoint}
(with respect to the sesquilinear form \eqref{form-1}) if
\[ e = e^\dagger. \]
In this case, the corresponding left and right projective modules
$\cM=\lAm\ast e$ and $\tilde\cM=e\ast\rAm$ are related by
\[ \tilde\cM=\left\{\zeta^\dagger \mid \zeta\in \cM \right\}.\]
Furthermore, the form \eqref{form-1} restricts to a sesquilinear
form on $\cM$, which is invariant under $\cG\cap U_m(\cA)$.

\begin{lemma}\label{lequalr}
Let $\cM=\lAm\ast e$ and $\tilde\cM=e\ast\rAm$ be the left and right
bundles associated with a self-adjoint idempotent $e$. Assume that
the left connection $\omega_i$ on $\cM$ and the right connection
$\tilde\omega_i$ on $\tilde\cM$ satisfy the condition
\begin{eqnarray}\label{omega-equality}
\tilde\omega_i = - \omega_i^\dagger, \quad \forall i.
\end{eqnarray}
Then for any $\zeta$ in $\cM$,
\[(\nabla_i\zeta)^\dagger = \tilde\nabla_i(\zeta^\dagger).\]
Furthermore, the curvatures on the left and right bundles are
related by
\[\tilde\cR_{i j} =- \cR_{i j}^\dagger.\]
\end{lemma}
\begin{proof}
Let $\xi=\zeta^\dagger$. We have
\[
(\nabla_i\zeta)^\dagger = (\partial_i\zeta + \zeta\ast \omega_i)^\dagger =
\partial_i\xi + \omega_i^\dagger \ast \xi = \tilde\nabla_i\xi.
\]
This proves the first part of the lemma. Now
\[
\begin{aligned}
\cR_{i j}^\dagger   &= \left(\partial_i \omega_j -\partial_j \omega_i
                    - [\omega_i, \ \omega_j]_\ast\right)^\dagger\\
                    & =\partial_i \omega_j^\dagger - \partial_j \omega_i^\dagger
                    + [\omega_i^\dagger, \ \omega_j^\dagger]_\ast
                    =-\tilde\cR_{i j}.
\end{aligned}
\]
This proves the second part.
\end{proof}

Hereafter we shall assume that condition \eqref{omega-equality} is
satisfied by the left and right connections. Let $\cM$ be the left
bundle corresponding to a self-adjoint idempotent $e$.  We shall say
that a connection $\omega_i$ on $\cM$ is {\em hermitian with respect
to the bar map} (or bar-hermitian) if $\omega_i^\dagger = \omega_i$
for all $i$. In this case, we shall also say that the bundle $\cM$
is {\em bar-hermitian}.

Note that the canonical connections $\omega_i = - \partial_i e$ on
$\cM$ and $\tilde\omega_i=\partial_i e$ on $\tilde\cM$ satisfy
$\tilde \omega_i = - \omega_i^\dagger$ and
$\omega_i^\dagger=\omega_i$ provided that $e$ is self-adjoint.
Therefore, in this case the canonical connection is bar-hermitian.
Since the left and right curvatures associated to the canonical
connections are equal, it follows from Lemma \ref{lequalr} that
$\cR_{i j}^\dagger = -\cR_{i j}$.

We have the following result.
\begin{theorem}\label{real}
Let $X=\begin{pmatrix} X^1 & X^2 &\dots & X^m\end{pmatrix}$ in
$\lAm$ be an embedded noncommutative surface satisfying the
condition $\overline{X}:=\begin{pmatrix} \overline{X^1} &
\overline{X^2} &\dots & \overline{X^m}\end{pmatrix}=X$.  Then $X$
has the following properties.
\begin{enumerate}
\item The metric has the property $\overline{g_{i j}}= g_{j i}$ for
all $i, j$.
\item The idempotent $e= (E_i)^t\ast g^{i j} \ast E_j$ is
self-adjoint.
\item Equipped with the canonical connection $\omega_i=-\partial_i
e$, the tangent bundle of $X$ is bar-hermitian.
\item The curvature satisfies $\cR_{i j}^\dagger = - \cR_{i j}$.
\end{enumerate}
\end{theorem}
\begin{proof}
The given condition on $X$ implies that all the $E_i$ satisfy
$E_i^\dagger = (E_i)^t$. Thus
\[ g_{i j} = E_i\ast (E_j)^t=E_i\ast
(E_j)^\dagger, \quad e=(E_i)^t\ast g^{i j} \ast E_j
=(E_i)^\dagger\ast g^{i j} \ast E_j.
\]
Hence we have
$
\overline{g_{i j}} = \left(E_i\ast
(E_j)^\dagger\right)^\dagger=E_j\ast (E_i)^\dagger = g_{j i}.
$
It then follows that $\overline{g^{i j}}=g^{j i}$. Now the
idempotent $e$ satisfies
\[
\begin{aligned}
e^\dagger &= \left((E_i)^\dagger\ast g^{i j} \ast E_j\right)^\dagger
=(E_j)^\dagger\ast \overline{g^{i j}} \ast E_i
&=(E_j)^t\ast g^{j i} \ast E_i =e.
\end{aligned}
\]
Part (3) and part (4) follow from part (2) and the discussion
preceding the proposition.
\end{proof}

Note that the quantum spacetimes studied in \cite{WZZ} and the
example in Section \ref{example} all satisfy the conditions of
Theorem \ref{real}.

\section{Concluding remarks}\label{conclusion}

We wish to point out that in the classical commutative setting, we
can recover (pseudo-) Riemannian geometry from the theory developed
here by using the isometric embedding theorems of \cite{embeddings,
embeddings-1, embeddings-2, embeddings-3}. The simplification in this case is that there is no
need to distinguish the left and the right tangent bundles. To
describe the situation, we let $(N, g)$ be a smooth $n$-dimensional
(pseudo-) Riemannian manifold with metric $g$. Denote by
$\cC^\infty(N)$ the set of smooth functions on $N$ endowed with the
usual pointwise multiplication. Let $\cC^\infty(N)^m$ be the space
consisting of row vectors of length $m$ with entries in
$\cC^\infty(N)$. By results of \cite{embeddings, embeddings-1, embeddings-2, embeddings-3},
there exist positive integers $p$, $q$ (with $p+q=m$) and a set of
smooth functions $X^1, \cdots, X^p, X^{p+1},\cdots, X^m$ on $N$ such
that
%\begin{eqnarray*}
$
g=\sum_{\alpha, \beta=1}^m dX^\alpha \eta_{\alpha \beta} dX^\beta,
$
%\end{eqnarray*}
where $\eta=diag(\underbrace{-1, \dots, -1}_p, \underbrace{1, \dots
1}_q)$ with $p=0$ if $N$ is Riemannian. Let $U$ be a coordinate
chart of $N$ with local coordinate $(t^1, \cdots, t^n)$. We set $E_i
=\begin{pmatrix} \frac{\partial X^1}{\partial t^i} & \frac{\partial
X^2}{\partial t^i}& \dots & \frac{\partial X^m}{\partial
t^i}\end{pmatrix}$ and define $e = \eta\left(E_i\right)^t g^{i j}
E_j$ on each coordinate chart $U$. Then we have the following
result.
\begin{theorem}\label{classical}
\begin{enumerate}
\item The idempotent $e$ is globally defined on $N$.
\item The space $\Gamma(TN)$ of sections of the tangent bundle of $N$ is given by
$\cC^\infty(N)^m e$.
\item For all $\zeta, \xi\in\Gamma(TN)$, we have $g(\zeta, \xi)=\zeta \eta (\xi)^t$.
\item The standard connection (with $\omega_i=-\partial_i e$) on $\cC^\infty(N)^m
e$ is the usual Levi-Civita connection on $TX$ with the Christoffel
symbol $\Gamma_{i j}^k$ defined by \eqref{Christoffel} and
$\Upsilon_{i j k}=0$.
\item The Riemannian curvature tensor is given by \eqref{R-tensor}.
\end{enumerate}
\end{theorem}

Returning to the noncommutative case, we recall that one can
quantise any Poisson manifold following the prescription of
\cite{Ko}. Then one obtains a collection of noncommutative
associative algebras (analogous to the Moyal algebra), one on each
coordinate patch. The algebras relative to different local
coordinates are gauge equivalent \cite[Theorem 2.3]{Ko} as discussed
in Section \ref{transformations}. This way, one obtains a sheaf of
noncommutative algebras over the Poisson manifold. The algebraic
geometry of such a quantised Poisson manifold has been extensively
developed by Kashiwara and Schapira \cite{KS1, KS2}. In principle one may
extend the local theory developed in this paper to a ``global"
differential geometry over the quantised Poisson manifold. Work in
this direction is currently under way.

Results in this paper should be directly applicable to the
development of a theory of noncommutative general relativity, which
is of considerable current interest in theoretical physics. We hope
that the theory presented here will provide a consistent
mathematical basis for this purpose. We should also mention that one
may use this theory to clarify, conceptually, aspects of the many
noncommutative geometries introduced in physics in recent years
based on physical intuitions. For example, general features of the
noncommutative geometries in \cite{Ch1, Ch2} and \cite{ADMW} have
considerable similarity with that of \cite{CTZZ}. These works also
have the advantage of being explicit and amenable to calculations,
thus have the chance to be physically tested. Therefore, it will be
useful to further develop the mathematical bases of these theories
by casting them into the framework of this paper.

Finally we note that a noncommutative analogue of spin geometry over
the Moyal algebra within the $C^*$-algebraic framework in terms of
noncompact spectral triples was studied in \cite{GBV}. Our treatment
is complementary to that of \cite{GBV}.

%

%\vspace{1cm}

\bigskip

\noindent{\bf Acknowledgement}: We wish to thank Masud Chaichian and
Anca Tureanu for discussions at various stages of this work. X.
Zhang thanks the School of Mathematics and Statistics, the
University of Sydney for the hospitality extended to him during a
visit when this work was completed. Partial financial support from
the Australian Research Council, National Science Foundation of
China (grants 10421001, 10725105, 10731080), NKBRPC (2006CB805905)
and the Chinese Academy of Sciences is gratefully acknowledged.

%

%\bigskip

\end{document}